\documentclass[sigconf, screen]{acmart}

\setcopyright{none}
\renewcommand\footnotetextcopyrightpermission[1]{}

\usepackage{textcomp}
\usepackage{array}
\usepackage{multirow}
\usepackage{xspace}
\usepackage{enumitem}

\newcommand{\circnum}[1]{\textcircled{\scriptsize #1}}

\newcommand{\shap}{\varphi}
\newcommand{\Lone}{$L_1$\xspace}
\newcommand{\hd}[1]{\textbf{\boldmath #1}}
\newcommand{\approach}{StepSplit\xspace}
\newcommand{\rulename}{step sharing\xspace}
\newcommand{\Rulename}{Step sharing\xspace}
\newcommand{\ctx}{\mathrm{ctx}}
\newcommand{\J}{\ensuremath{\,\mathrm{J}}\xspace}

\begin{document}

\title{Scaling Shapley Energy Attribution to Large LLM Batches}

\author{Qi Luo}
\email{qluo615@connect.hkust-gz.edu.cn}
\affiliation{
  \institution{The Hong Kong University of Science and Technology (Guangzhou)}
  \city{Guangzhou}
  \country{China}}

\author{Ziwen Wang}
\email{zwang578@connect.hkust-gz.edu.cn}
\affiliation{
  \institution{The Hong Kong University of Science and Technology (Guangzhou)}
  \city{Guangzhou}
  \country{China}}

\author{Kunlin Li}
\email{kli082@connect.hkust-gz.edu.cn}
\affiliation{
  \institution{The Hong Kong University of Science and Technology (Guangzhou)}
  \city{Guangzhou}
  \country{China}}

\author{Changxi Liu}
\email{changxiliu@hkust-gz.edu.cn}
\affiliation{
  \institution{The Hong Kong University of Science and Technology (Guangzhou)}
  \city{Guangzhou}
  \country{China}}

\author{Yun Chen}
\authornote{Corresponding author.}
\email{yunchen@hkust-gz.edu.cn}
\affiliation{
  \institution{The Hong Kong University of Science and Technology (Guangzhou)}
  \city{Guangzhou}
  \country{China}}

\begin{abstract}
LLM services have become essential infrastructure for a wide range of applications, such as chat assistants and coding agents. Usually, service providers need to attribute resource consumption to individual requests for purposes such as billing. Among these resources, energy is a key metric for assessing the cost of serving each request. However, serving engines batch requests together, and GPUs report only aggregate energy consumption, making it difficult to attribute energy use to individual requests. Providers typically estimate each request's share in proportion to its token consumption, but this approach can produce substantially unfair estimates, especially when requests vary significantly in length or computational complexity. The Shapley value provides a principled solution to fair energy allocation, but computing it directly requires $2^N-1$ GPU runs for $N$ requests, which is infeasible for large batches.

To address this problem, we propose \textbf{\approach}, an efficient method for attributing energy consumption to individual LLM requests. Across GPU generations, we observe that the energy consumed by each inference step, a forward pass over a batch, consists of a shared energy, largely due to reading model weights, and a request-specific energy. Based on this observation, we derive a closed-form Shapley value: each request is charged its own energy plus an equal part of the shared energy of every step in which it participates. A one-time microbenchmark lasting 2.5--17.8 minutes calibrates both parts of the step energy for each model and GPU configuration. Under static batching, \approach uses this closed form in a single pass over the scheduler log, which records the requests served at each step. Under continuous batching, removing a request can change how the remaining requests are batched, so \approach adds a correction to the closed form, computed by replaying the scheduler on a CPU. We evaluate \approach against Shapley values measured using 529{,}513 GPU runs. The results show that, under static batching with batch sizes of 8--256, \approach achieves an \Lone error of 0.4--4.3\%, compared with 13.4--61.3\% for token-proportional allocation. Under continuous batching with 8--128 requests, the corresponding errors are 0.5--6.4\% and 20.6--71.7\%.

\end{abstract}

\maketitle
\pagestyle{plain}

\section{Introduction}
\label{sec:intro}

LLM services have become essential infrastructure for many applications, such as chat assistants and coding agents~\cite{patel2024splitwise,wang2024burstgpt,qin2025mooncake}. Service providers need to attribute resource consumption to individual requests for purposes such as billing~\cite{neuralwatt2026methodology}, carbon reporting~\cite{elsworth2025google,rince2025ecologits} and scheduling~\cite{shen2013power}. Among these resources, energy is a key metric for assessing the cost of serving each request, as it is a major cost of LLM inference~\cite{oviedo2026energy,patel2026frontiers}. However, the energy of a single request cannot be measured directly~\cite{roy2026identifiability}. A serving engine such as vLLM~\cite{kwon2023vllm} runs many requests together in each forward pass. The GPU draws approximately 75\% of the power of a server~\cite{dodge2022measuring,patel2024characterizing,latif2025single,aquinobritez2025towards} and reports only their total energy. How to split this total fairly among the requests is not obvious.

\begin{figure}[t]
\centering
\includegraphics[width=0.92\linewidth]{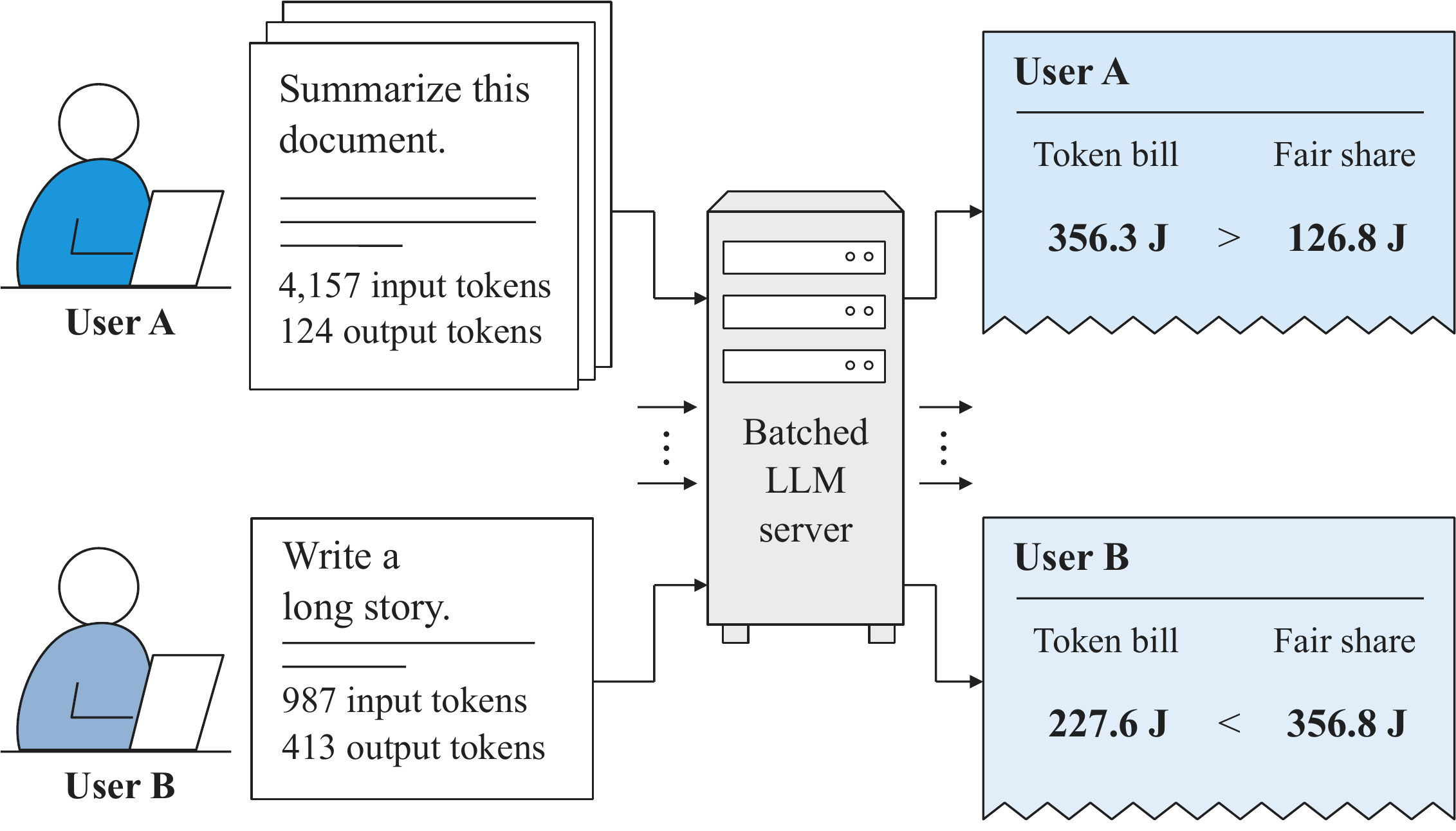}
\caption{Comparison of token bills and fair Shapley shares.}
\label{fig:teaser}
\end{figure}

In practice, the cost of a request is usually measured by its tokens. OpenAI and Anthropic price requests per token~\cite{openai2026pricing,anthropic2026pricing}, a commercial service bills energy by the share of each request in the in-flight tokens~\cite{neuralwatt2026methodology}, and fair schedulers weight requests by their tokens~\cite{sheng2024fairness}. As shown in Figure~\ref{fig:teaser}, splitting by tokens does not give each request its fair share. The reason is that each forward pass reads the weights once, however many requests it holds. As we show in \S\ref{sec:char}, the energy of a pass is the shared energy of this read plus the own energy of each request. The read takes 11.0--89.8\% of the energy, depending on the workload, and all requests in the pass share it. The own energy of a request grows with its tokens, but its share of the reads depends on how many passes it joins and how many requests share each pass, not on its tokens.

The standard fair split is the Shapley value~\cite{shapley1953value}. It charges each request the extra energy it causes, averaged over every order in which the requests could join. Energy and carbon attribution use it as their reference~\cite{dong2014rethink,han2025fairco2,li2025fair}. However, computing it requires $2^N-1$ separate GPU runs for the subsets of the $N$ requests, 255 for 8 requests but $4.3\times10^{9}$ for 32, which is infeasible for large batches. In practice, it is measured only for small groups such as 8 requests~\cite{roy2026seam,luo2026request}. Scaling to larger groups needs a sampled estimate~\cite{castro2009polynomial}, whose number of runs still grows with the group. Yet a serving engine batches hundreds of requests, so neither approach reaches the batch sizes of real serving. Prior work instead approximates the value with a regression trained on measured values~\cite{luo2026request}, a game over time periods rather than requests~\cite{han2025fairco2}, or Shapley-inspired splits of idle energy~\cite{sharma2024faasmeter}. Providers thus have a scalable split that is not fair and a fair split that does not scale.

To address this problem, we propose \textbf{\approach}. We call each forward pass a \emph{step}. We measure the two-part energy on three GPU generations, and the shared energy of a step stays the same however many requests are in it. The energy of any subset can therefore be computed from its steps instead of measured in a separate GPU run. Under this structure, the Shapley value has a closed form known from classic cost-sharing problems~\cite{littlechild1973simple,feigenbaum2001sharing}. Each request pays its own energy plus an equal part of the shared energy of every step it joins. We call this rule \emph{\rulename}. A microbenchmark of 2.5--17.8 minutes calibrates both parts of the step energy once per model and GPU configuration.

Under static batching, \approach uses this closed form in one pass over the scheduler log of the run that serves the requests, without enumerating subsets. Under continuous batching, removing a request changes when the other requests run, so a subset goes through different steps from those in the scheduler log, and the closed form on the logged steps no longer gives the Shapley value. \approach therefore adds a correction to the closed form, computed by replaying the scheduler of the serving engine on a CPU. The closed form does not give this correction, so \approach samples it~\cite{castro2009polynomial}, and for each sampled subset it computes the energy of the replayed and of the logged steps on the CPU instead of measuring it on the GPU. We also build an online ledger that hooks into vLLM without changing it. As each step ends, the ledger adds to every scheduled request its share of the step, at 47\,$\mu$s per step, so each request is billed when it finishes. Under continuous batching, more accurate bills need a CPU replay after the run.

We compare \approach with Shapley values (exact for 8 requests, sampled for more) measured on the GPU across 529{,}513 runs, seven models, three GPU generations and a distributed deployment. Under static batching with batch sizes of 8--256, \approach has an \Lone error of 0.4--4.3\%, against 13.4--61.3\% for splitting by tokens and 8.6--26.6\% for a regression trained on measured Shapley values~\cite{luo2026request}. For 8--128 requests under continuous batching, it has 0.5--6.4\%, against 20.6--71.7\% and 8.0--56.0\%.

In summary, we make the following contributions:
\begin{itemize}[leftmargin=*,labelindent=0pt,nosep]
\item We characterize the energy of batched LLM inference on three GPU generations and show that the energy of each step has two parts: a shared energy, mostly for reading the model weights, and the own energy that each request adds.
\item We propose \approach, which computes the Shapley value of each request in one pass over the scheduler log of one run, with no extra GPU runs. Under continuous batching and in a distributed deployment, it replays the scheduler and the router on a CPU. The same rule extends to MoE experts and cached prefixes.
\item We validate \approach against Shapley values measured in 529{,}513 GPU runs and show that it is more accurate than both splitting by tokens and a trained regression in every configuration.
\item We build an online ledger that hooks into vLLM without changing it and bills each request when it finishes, at 47\,$\mu$s per step.
\end{itemize}

\section{Background}
\label{sec:background}

\subsection{Cooperative Games and the Shapley Value}
\label{sec:game}

A fair split charges each request for the energy it causes. However, requests served together share part of the work, so the energy that a request adds depends on which other requests are served with it. Cooperative game theory defines this split with the Shapley value~\cite{shapley1953value}. The players are the $N$ requests $\mathcal{N}$ served together. For a coalition $S\subseteq\mathcal{N}$, the characteristic function $v(S)$ is the energy of serving only the requests in $S$, with $v(\emptyset)=0$. The Shapley value of request $i$ averages the energy that $i$ adds to the requests before it over all orders in which the requests could join,
\begin{equation}
\shap_i \;=\; \frac{1}{N!}\sum_{\sigma}\Bigl[v\bigl(P_i^{\sigma}\cup\{i\}\bigr)-v\bigl(P_i^{\sigma}\bigr)\Bigr],
\label{eq:shapley}
\end{equation}
where $P_i^{\sigma}$ is the set of requests that precede $i$ in the order $\sigma$. It is the only split that satisfies four axioms. Its shares add up to $v(\mathcal{N})$ (efficiency), treat interchangeable requests equally (symmetry), give nothing to a request that never adds energy (null player), and add up over games (additivity). Since each $v(S)$ is a separate GPU run, the definition needs $2^N-1$ runs, 255 for 8 requests but about $4.3\times10^{9}$ for 32.

\subsection{Batched LLM Inference}
\label{sec:inference}

A serving engine such as vLLM~\cite{kwon2023vllm} runs an LLM on GPUs and serves a batch of requests one step at a time. Each \emph{step} is one forward pass of the model. A request $i$ has a prompt of $p_i$ input tokens and generates $d_i$ output tokens. Each request goes through two phases, \emph{prefill} and \emph{decode}. Prefill processes the prompt and stores the result in a KV cache. Decode then generates one output token per step and reads the KV cache each time. A step handles all of its requests together, whether in prefill or in decode. In every step the GPU reads the model weights from its memory once, for all of its requests. This read is the work that the requests of a step share. \emph{Static batching} serves requests that are all available at the start. They enter the batch together, so $N$ is the batch size, and the number of requests in a step, $m$, starts at $N$ and falls as requests finish. \emph{Continuous batching}~\cite{yu2022orca} serves requests that arrive over time, as in online serving, and lets them join and leave the running batch between steps, so $m$ changes from step to step.

\subsection{The Energy of a Coalition}
\label{sec:energy}

We split the \emph{busy energy} of the GPUs, which is the total energy that they consume while they serve requests, including the \emph{idle power} they draw with the model loaded even without work. The gaps without any request, such as while the engine waits for arrivals, and host stalls (\S\ref{sec:setup}) are left out, because no request causes their energy. The energy law of \S\ref{sec:theory} includes this idle power in its coefficients. In our experiments on three GPU generations, the idle power contributes 24.4--46.2\% of the busy energy.

Defining $v(S)$ also needs the number of tokens that each request generates in a coalition. Batched arithmetic is not bit-identical across batch compositions, so even with greedy decoding the output length of a request can change with the other requests. If output lengths were left free, the energy a request adds would include the energy of length changes in other requests, which it did not cause. We therefore define $v(S)$ with every request generating the same number of tokens as in the actual run, which is also the number an online attribution system observes.

\section{Motivation}
\label{sec:motivation}\label{sec:example}

\begin{table}[ht]
\caption{Input and output tokens of 8 requests in one batch and the energy (J) attributed to each.}
\label{tab:example}
\centering
\begin{small}
\setlength{\tabcolsep}{4.2pt}
\begin{tabular}{rrrrrrr}
\toprule
\hd{Request} & \hd{$p_i$} & \hd{$d_i$} & \hd{Token} & \hd{Solo} & \hd{JCalib} & \hd{Shapley} \\
\midrule
0 & 120 & 14 & 14.2 & 11.4 & 21.3 & 7.6 \\
1 & 1060 & 397 & 227.1 & 304.5 & 335.1 & 315.8 \\
2 & 987 & 413 & 227.6 & 315.5 & 351.4 & 356.8 \\
3 & 1314 & 176 & 163.6 & 138.5 & 103.1 & 113.9 \\
4 & 4157 & 124 & 356.3 & 112.8 & 135.3 & 126.8 \\
5 & 173 & 110 & 53.9 & 83.4 & 31.2 & 50.2 \\
6 & 1232 & 399 & 240.7 & 306.7 & 341.8 & 323.5 \\
7 & 204 & 67 & 40.2 & 50.9 & 4.3 & 29.0 \\
\midrule
\multicolumn{3}{l}{\Lone vs.\ Shapley} & 0.454 & 0.126 & 0.090 & -- \\
\bottomrule
\end{tabular}
\end{small}
\end{table}

Because the exact Shapley value is expensive, the energy of a batch is often split with cheaper rules. \emph{Token} splits it in proportion to the tokens of each request. Following the pricing of Anthropic and OpenAI~\cite{anthropic2026pricing,openai2026pricing}, we count the tokens of request $i$ as $p_i+5d_i$. \emph{Solo} splits it in proportion to the energy of each request run alone. \emph{JCalib}~\cite{luo2026request} is a regression that predicts shares from features of the token counts of each request and its batch, trained on the measured Shapley values of other batches. Table~\ref{tab:example} compares these rules with the exact Shapley value of one batch of 8 requests from the Azure conversation trace~\cite{patel2024splitwise}, measured with all 255 runs of Qwen2.5-7B on an H100 under static batching. We measure the error of a rule by the normalized \Lone distance $\sum_i|\hat\shap_i-\shap_i|/v(\mathcal{N})$ between its shares $\hat\shap_i$ and the Shapley values $\shap_i$ (Eq.~\eqref{eq:shapley}).

Token has the largest \Lone error, 45.4\%. It overcharges requests with many input tokens and undercharges those with many output tokens. Request~4, with 4{,}157 input and 124 output tokens, is charged 356.3\J against a Shapley share of 126.8\J. Solo is closer at 12.6\%, but it needs one extra GPU run per request, which is impractical in serving. It charges the short requests 0, 5 and~7 $1.5$--$1.8\times$ their shares, because alone each pays for the weight reads of all its steps, which it shares with most of the other requests in the batch. JCalib comes closest at 9.0\%. However, its features describe token counts, not how many requests share each step. All three rules miss the \emph{shared energy} that each step consumes once for all its requests, mainly to read the weights. In this batch, the shared energy of its 413 steps is 83.7\% of the total. The fair shares thus depend mainly on which steps each request is in, and the output lengths set these steps, not the input lengths.

\begin{figure}[h]
\centering
\includegraphics[width=0.92\linewidth]{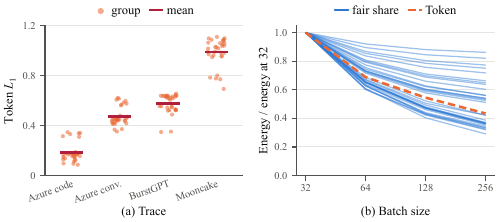}
\caption{(a)~Token \Lone for each group of 8 requests, by trace. (b)~Fair share of each of the same 32 requests as the batch grows, relative to a batch of 32.}
\label{fig:billing}
\end{figure}

The error of Token is not limited to this batch. As shown in Figure~\ref{fig:billing}a, it appears in every group of 8 requests that we measure, and it differs across workloads. The mean Token \Lone is 18.5\% on the Azure code trace and 98.7\% on the Mooncake agent trace~\cite{qin2025mooncake}. As shown in Figure~\ref{fig:billing}b, the fair share of a request also depends on the batch it runs in. We draw 256 requests from the Azure conversation trace and serve the first 32 of them in a batch of 32 and then in batches of 64 to 256. In the batch of 256, their fair shares fall to 0.29--0.86 of those in the batch of 32. We compute these fair shares with the method of \S\ref{sec:method}, which matches the sampled Shapley values of these batches (\S\ref{sec:eval:large}). In the larger batch, more requests share each step, so each request pays a smaller part of the shared energy of the step. Requests with long outputs, whose energy is mostly shared, thus fall the most. Token cannot follow these changes. It lowers the charge of every one of the 32 requests to the same 0.43 of its charge in the batch of 32, because only the energy per token of the batch changes, not the tokens of the request.

None of the three rules gives fair shares. Computing the Shapley value without running every subset on the GPU thus needs an analytical model of the energy of a subset. \S\ref{sec:char} characterizes how the energy of a step depends on the requests in it, and \S\ref{sec:method} builds this model from it.

\section{Characterizing the Energy of a Step}
\label{sec:char}

\begin{figure*}[t]
\centering
\includegraphics[width=0.92\linewidth]{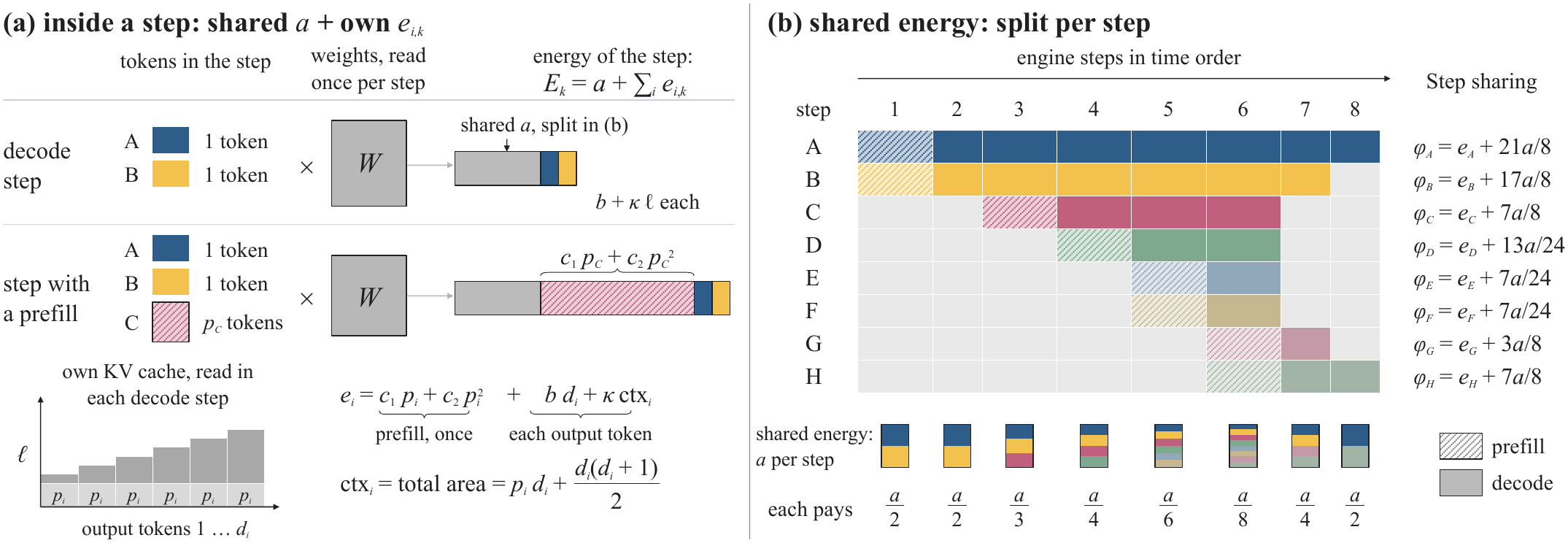}
\caption{\Rulename, $\shap_i = e_i + s_i$. (a)~Inside step $k$, each scheduled request $i$ adds its own energy $e_{i,k}$, while the step incurs the shared energy $a$ once. (b)~The shared energy $a$ of each step is split equally among the requests in it, however long the step lasts.}
\label{fig:overview}
\end{figure*}

Section~\ref{sec:platform} gives the setup, which \S\ref{sec:eval} also uses. Section~\ref{sec:insights} presents five insights that guide the design of \approach.

\subsection{Experimental Setup}
\label{sec:platform}

\textbf{Platforms and models.} We evaluate three NVIDIA GPU generations: Hopper H100, Ampere A800, and Blackwell RTX PRO 6000 (PRO 6000). All experiments run on vLLM~0.30 with greedy decoding, driven one step at a time by a harness that logs the duration of every step and the tokens of every scheduled request. We test seven models spanning three architectures: the dense Qwen2.5-7B, Qwen2.5-14B, and Qwen2.5-72B~\cite{qwen2025qwen25} and Gemma-4-12B~\cite{gemma2026gemma4}; the MoE models gpt-oss-20b~\cite{openai2025gptoss} and Qwen3-30B-A3B~\cite{yang2025qwen3}; and the linear-attention hybrid Qwen3.8-27B~\cite{qwen2026qwen38}. Models run on a single GPU or with tensor parallelism across two or four GPUs at default clocks and power limits. Weights are in BF16, except for the MXFP4 experts of gpt-oss-20b. We also evaluate a \emph{distributed deployment} serving Qwen2.5-72B across two nodes behind a router that dispatches each arriving request to the less loaded node, where each node runs one vLLM engine on four H100s.

\textbf{Workloads.} We take request lengths and arrival times from four production traces: the Azure LLM inference traces for conversation and code~\cite{patel2024splitwise}, BurstGPT~\cite{wang2024burstgpt}, and the Mooncake agent trace~\cite{qin2025mooncake}. Because these traces omit prompt text, we truncate ShareGPT~\cite{sharegpt} conversations to the input length of each request and fix each output to its trace length. We run 8 to 256 requests under static batching and 8 to 128 requests under continuous batching. We measure the exact Shapley value over all subsets for 8 requests and estimate it by sampling for 16 or more requests (\S\ref{sec:setup}). Under continuous batching, the 16 to 128 requests form a \emph{trace window} of consecutive requests that arrive at their recorded times. A single engine serves these windows, and the distributed deployment also serves trace windows of 32 to 128 requests. At their recorded rates, the conversation and agent traces average 7--10 requests per step. To reach larger batches, we also replay them at $3\times$ and $10\times$ speed, keeping the request lengths and the burst pattern of the arrivals.

\textbf{Microbenchmark.} A synthetic microbenchmark serves $m$ requests with input length $p$ and output length $D$ under static batching, so that every decode step holds exactly $m$ requests and the three vary independently. Each input is a different stretch of $p$ tokens from the ShareGPT conversations. The grid is $m\in\{1,2,4,\dots,256\}$, $p\in\{128,512,2048,8192\}$ with $mp\le131{,}072$, and $D\in\{1,64,256\}$, with two runs for each combination, 180 runs in all. Runs whose KV cache does not fit in GPU memory are skipped. It runs once per model and GPU configuration (GPU model and number of GPUs).

\textbf{Measuring energy.}\label{sec:measure}\label{app:protocols} We read the cumulative energy counter of each GPU through NVML~\cite{nvml}, which updates at 10\,Hz, and sum the energies of all GPUs that run the model. For the busy energy of \S\ref{sec:energy}, we subtract the idle power over the time outside steps. We measure the idle power of each GPU with the model loaded, before its runs.

\subsection{Insights}
\label{sec:insights}

\begin{figure}[h]
\centering
\includegraphics[width=0.92\linewidth]{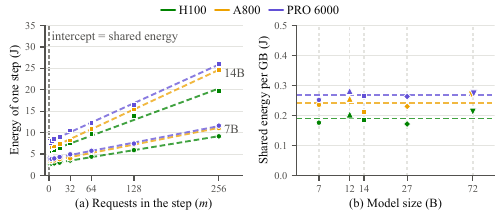}
\caption{(a)~Energy of one decode step against the number of requests in it, with fitted lines. (b)~Shared energy per GB of weights against model size (log axis). Each dashed line is the average of the points of one GPU.}
\label{fig:step}
\end{figure}

\textbf{Insight 1. The energy of a step is a shared energy plus the own energy of each request.} The shared energy is paid once per step, and the own energy grows with the tokens of the request (Figure~\ref{fig:overview}a). We call this decomposition the \emph{energy law} and give its terms in \S\ref{sec:theory}. In the microbenchmark, for the same $m$ and $p$, a longer output $D$ only adds decode steps of $m$ requests, so the extra energy per added step is the energy of one decode step. It grows nearly linearly with $m$, from 2.66\J with one request to 9.15\J with 256 at input length 512 for Qwen2.5-7B on the H100. The intercept of the fitted line is the shared energy and its slope is what one more request adds. Qwen2.5-14B, the A800 and the PRO 6000 behave the same way (Figure~\ref{fig:step}a). Calibrated on the microbenchmark alone, the law also predicts the energy of subsets of static batches from the Azure conversation and BurstGPT traces on the H100, with an average error of 0.6--2.8\% for Qwen2.5-7B and 0.8--3.8\% for Qwen2.5-14B. The energy of a coalition is therefore set by its steps and the tokens it processes in them.

\begin{figure}[h]
\centering
\includegraphics[width=0.92\linewidth]{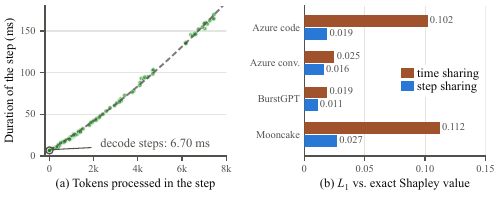}
\caption{(a)~Duration of each step against the tokens it processes. (b)~Error of splitting the shared energy by time or by step.}
\label{fig:time}
\end{figure}

\textbf{Insight 2. The shared energy is mostly the energy of reading the weights once per step.} A decode step of Qwen2.5-7B on the H100 lasts 6.70\,ms at the median, and the fitted curve of Figure~\ref{fig:time}a meets zero tokens at 6.68\,ms, so most of a step is the work that it does once for all of its requests, which is reading the weights. If the shared energy is the energy of this read, it grows in proportion to the size of the weights, at a fixed energy per GB on each GPU. As shown in Figure~\ref{fig:step}b, the shared energy per GB of weights stays nearly the same on each GPU from 7B to 72B, at 0.172--0.212\J on the H100, 0.211--0.270\J on the A800 and 0.252--0.282\J on the PRO 6000, including Qwen2.5-72B on four GPUs (two on the PRO 6000). The shared energy is therefore set by the model and the GPU, and a microbenchmark can calibrate it once per model and GPU configuration.

\textbf{Insight 3. The shared energy is paid per step, not per unit of time.}\label{sec:time}\label{sec:eval:unit} Time sharing divides static, idle or shared chip power equally among the tenants or tasks active at each instant~\cite{islam2016new,shen2013power,sharma2024faasmeter}. This fits the shared energy only if it grows with the duration of a step, but every step reads the weights once however long it lasts. As shown in Figure~\ref{fig:time}a, the rest of a step grows with its tokens, so a long step is long because of its own tokens. The idle power drawn over this extra time therefore also belongs to the requests that hold the tokens. In the measured run that Figure~\ref{fig:overview} sketches, a step that prefills a 7{,}142-token input lasts 158\,ms, which is $23.6\times$ the 6.70\,ms of a decode step, so time sharing bills its shared energy as $23.6a$ instead of $a$. It charges $A$ and $B$, which only decode, $7.9a$ each for the prefill of $C$ (step~3 of Figure~\ref{fig:overview}b). Against the exact Shapley values of the 8-request groups under continuous batching, with the same own energies, time sharing has an \Lone error of 0.019--0.112 and step sharing 0.011--0.027 (Figure~\ref{fig:time}b), with the largest gap on the long inputs of the code and Mooncake traces.

\begin{figure}[h]
\centering
\includegraphics[width=0.92\linewidth]{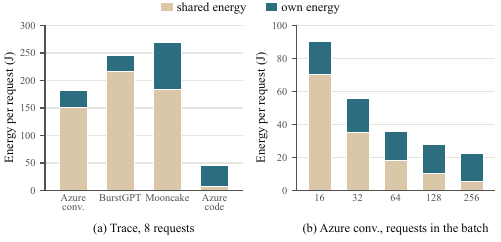}
\caption{(a)~Energy per request in static 8-request batches of each trace. (b)~Energy per request against batch size in static batches from the Azure conv. trace.}
\label{fig:reqenergy}
\end{figure}

\textbf{Insight 4. The fraction of shared energy varies widely across traces and batch sizes.} Because the shared energy of a step is fixed while the own energy grows with the tokens in it, the fraction of shared energy depends on the lengths of the requests and on how many of them share each step. For Qwen2.5-7B on the H100 it is 0.110--0.898 of the batch energy. As shown in Figure~\ref{fig:reqenergy}a, in 8-request batches the shared energy is 0.832 and 0.886 on the two conversation traces and 0.681 on the agent trace, but only 0.183 on the code trace, where long inputs and short outputs make the own energy dominate. As shown in Figure~\ref{fig:reqenergy}b, it also falls as more requests share the steps. In static batches drawn from the Azure conversation trace, the shared energy of a request falls from 70.6\J with 16 requests to 5.4\J with 256, while its own energy stays at 16.9--20.5\J. No single price per token can therefore track the shared energy, and a fair split has to follow the steps of the batch.

\textbf{Insight 5. Under continuous batching and distributed deployment, the steps of a request depend on the other requests.} Removing requests shortens the remaining steps, and later arrivals join different ones. On 40 random subsets of each of seven 128-request trace windows, the law predicts the energy of a subset with an average error of 0.2--0.7\% when given the steps that the subset really takes, but 1.4--5.7\% when given the steps of the full run with the other requests removed. In the distributed deployment, 41--50\% of the pairs of requests change whether they share a node. The steps of a coalition are therefore not in the log of the full run and have to be reconstructed.

\section{\approach}
\label{sec:method}\label{sec:system}

\begin{figure}[t]
\centering
\includegraphics[width=0.92\linewidth]{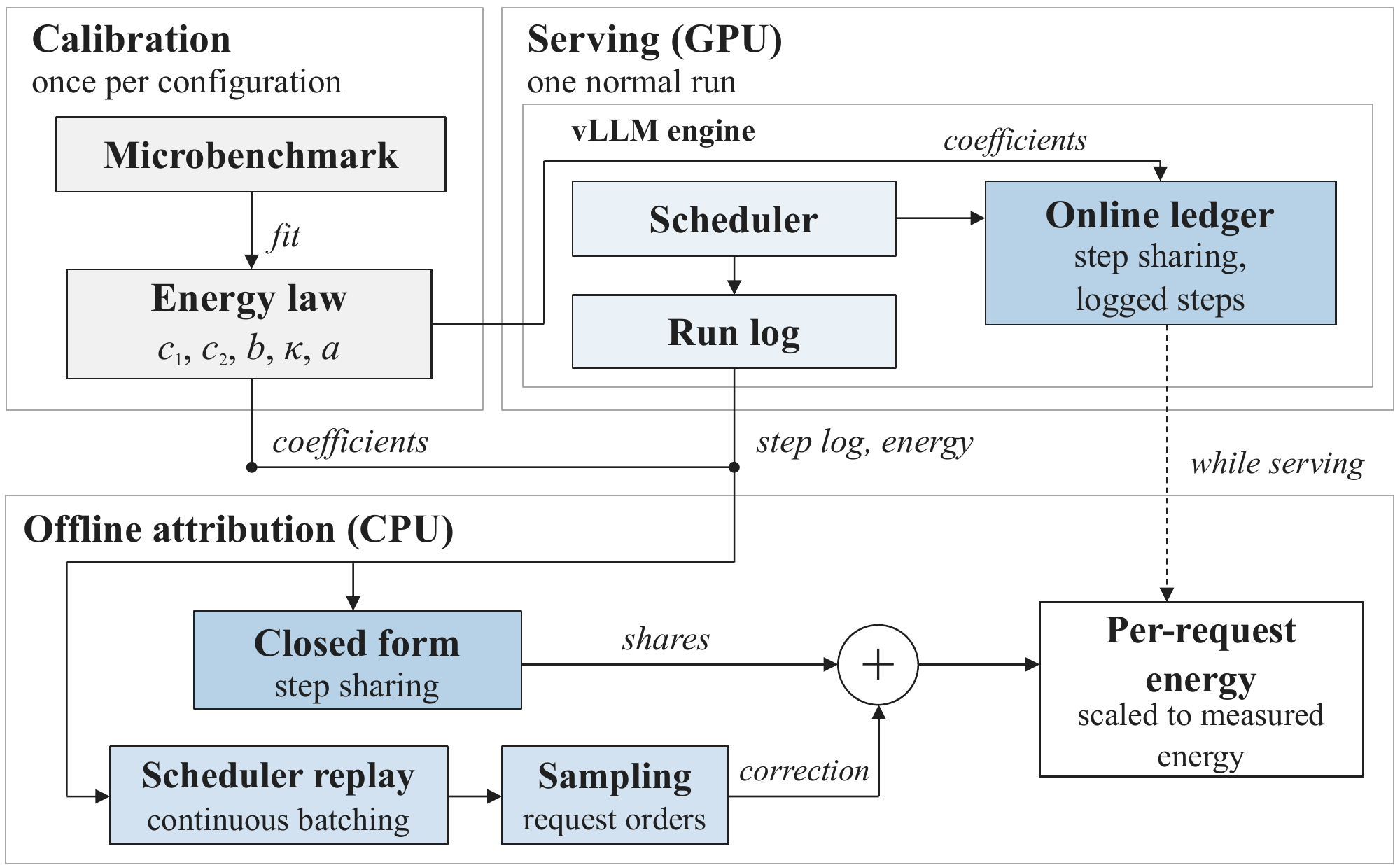}
\caption{Overview of \approach.}
\label{fig:arch}
\end{figure}

As shown in Figure~\ref{fig:arch}, \approach turns one run of the attributed workload into per-request shares. A microbenchmark calibrates the energy law once per model and GPU configuration. While the engine runs, an online ledger computes the closed-form shares from the scheduler log. After a run under continuous batching, offline attribution replays the scheduler on a CPU and adds the resulting correction to the closed-form shares. Both divide the measured energy among the requests in proportion to these shares.

\subsection{A Closed-Form Shapley Value}
\label{sec:theory}

Following Insight~1, let a run consist of steps $k=1,\dots,K$. Let $M_k\subseteq\mathcal{N}$ be the $m_k$ requests scheduled in step $k$, whether prefilling or decoding. Engines cap the tokens per step, so a long input is prefilled in chunks over several steps alongside the decode tokens of other requests (\emph{chunked prefill}~\cite{agrawal2024sarathi}). We model the energy of step $k$ as
\begin{equation}
E_k \;=\; a\;+\;\sum_{i\in M_k} e_{i,k},
\label{eq:law}
\end{equation}
where the shared energy $a$ is the same in every step and $e_{i,k}$ depends only on the work of request $i$ in the step. A chunk of $\pi_{i,k}$ input tokens starting at input position $o_{i,k}$ costs $c_1\,\pi_{i,k}+c_2\bigl[(o_{i,k}+\pi_{i,k})^2-o_{i,k}^2\bigr]$, and an output token produced at context length $\ell_{i,k}$ costs $b+\kappa\,\ell_{i,k}$.

Here $c_1$ is the computation per input token, $c_2$ is the attention among the input tokens, $b$ is the computation per output token, $\kappa$ is the read of one cached token, and $a$ is mostly the read of the weights. Idle power enters each coefficient through the time its work takes. Summed over the steps of a request, these terms give its \emph{own energy}
\begin{equation}
e_i = c_1\,p_i + c_2\,p_i^2 + b\,d_i + \kappa\,\ctx_i,
\quad \ctx_i = \sum_{j=1}^{d_i}\,(p_i+j).
\label{eq:own}
\end{equation}
The step that ends the prompt also produces the first output token, so a request produces one token in each of $d_i$ steps. The prefill terms telescope over chunks, so chunking does not change $e_i$.

The steps in the log of the full run form its \emph{logged timeline}. On this timeline, the energy law gives every coalition $S$ the energy
\begin{equation}
\hat v(S) \;=\; \sum_{i\in S} e_i \;+\; a\,K(S),
\label{eq:game}
\end{equation}
where $K(S)=\bigl|\{k:\,M_k\cap S\neq\emptyset\}\bigr|$ is the number of logged steps that hold a member of $S$. We call $\hat v$ the \emph{model game} to tell it from the measured game $v$ of \S\ref{sec:game}. The output length $d_i$ of each request is fixed (\S\ref{sec:energy}). Under static batching with every prompt prefilled in the first step, request $i$ is therefore scheduled in steps 1 to $d_i$ whatever else is served. The logged steps that hold a member of $S$ are then exactly the steps that $S$ goes through when served alone, and step $k$ holds $M_k\cap S$. Under continuous batching, the steps of a coalition differ from the logged ones, and \S\ref{sec:replay} reconstructs them.

\begin{proposition}[\cite{littlechild1973simple,ginsburgh2003museum}]
\label{thm:main}
The Shapley value of request $i$ in the model game is exactly
\begin{equation}
\shap_i \;=\; e_i \;+\; s_i,\qquad s_i \;=\; \sum_{k:\,i\in M_k} \frac{a}{m_k}.
\label{eq:rule}
\end{equation}
This result holds unchanged when the shared energy of a step depends on the number of requests in it, with $a(m_k)$ in place of~$a$.
\end{proposition}

\begin{proof}
With $a(0)=0$, write the model game as $\hat v=v_0+\sum_{k}v_k$ with $v_0(S)=\sum_{i\in S}e_i$ and $v_k(S)=a(|M_k\cap S|)$. The Shapley value is linear in the game. In $v_0$ every marginal contribution of $i$ is $e_i$, so its share is $e_i$. In $v_k$ a request outside $M_k$ never changes the value and receives nothing. Two requests in $M_k$ change $|M_k\cap S|$ in the same way for every $S$ that contains neither. They therefore receive equal shares, which add up to $v_k(\mathcal{N})=a(m_k)$, so each request in $M_k$ receives $a(m_k)/m_k$. Summing over $k$ gives Eq.~\eqref{eq:rule}. No monotonicity or concavity of $a(m)$ is used.
\end{proof}

We call Eq.~\eqref{eq:rule} \rulename. As shown in Figure~\ref{fig:overview}b, a request pays its own energy $e_i$ and its \emph{shared energy} $s_i$, which is an equal part of the shared energy of every step it is in. The rule replaces the $2^N-1$ runs of Eq.~\eqref{eq:shapley} by one pass over the log of the run being attributed in $O(\sum_k m_k)$ operations.

The result is classical, and the new finding is that measured GPU energy has this structure. Under static batching, the shared term of Eq.~\eqref{eq:game} is $a\max_{i\in S}d_i$, which is the airport game~\cite{littlechild1973simple}. On the logged steps of continuous batching, Eq.~\eqref{eq:game} is a coverage game with the same equal-split solution~\cite{feigenbaum2001sharing,ginsburgh2003museum}.

\subsection{Calibration}
\label{sec:calib}\label{app:calib}

The microbenchmark of \S\ref{sec:platform} calibrates the law without trace data. A configuration that serves only the 8-request groups of \S\ref{sec:eval:exact} runs the 96 runs with $m\le8$, which cover the requests per step it serves. A configuration that also serves larger batches runs all 180. For Qwen2.5-7B on the H100, the 180 runs take 7.0 minutes, and the runs of every configuration take 2.5--17.8 minutes. Non-negative least squares fits the five coefficients from these runs, and every experiment of a configuration and its online ledger use them. Cards of the same GPU model differ, and the fitted shared energy of two H100s differs by 3.0--3.7\%. To keep this difference out of the validation, we fit the coefficients for each card when the 8-request groups of a configuration ran on several cards. In a tensor-parallel group, the GPUs execute every step together, so the run keeps one step log. Their summed energy follows the same law with its own coefficients, and the shared energy then also covers the communication between the GPUs. Because the measured energy is divided in proportion to the shares, a common scale error in the coefficients cancels and only the ratios $c_1:c_2:b:\kappa:a$ matter.

\subsection{Scheduler Replay}
\label{sec:replay}\label{sec:drift}

\begin{figure}[t]
\centering
\includegraphics[width=0.92\linewidth]{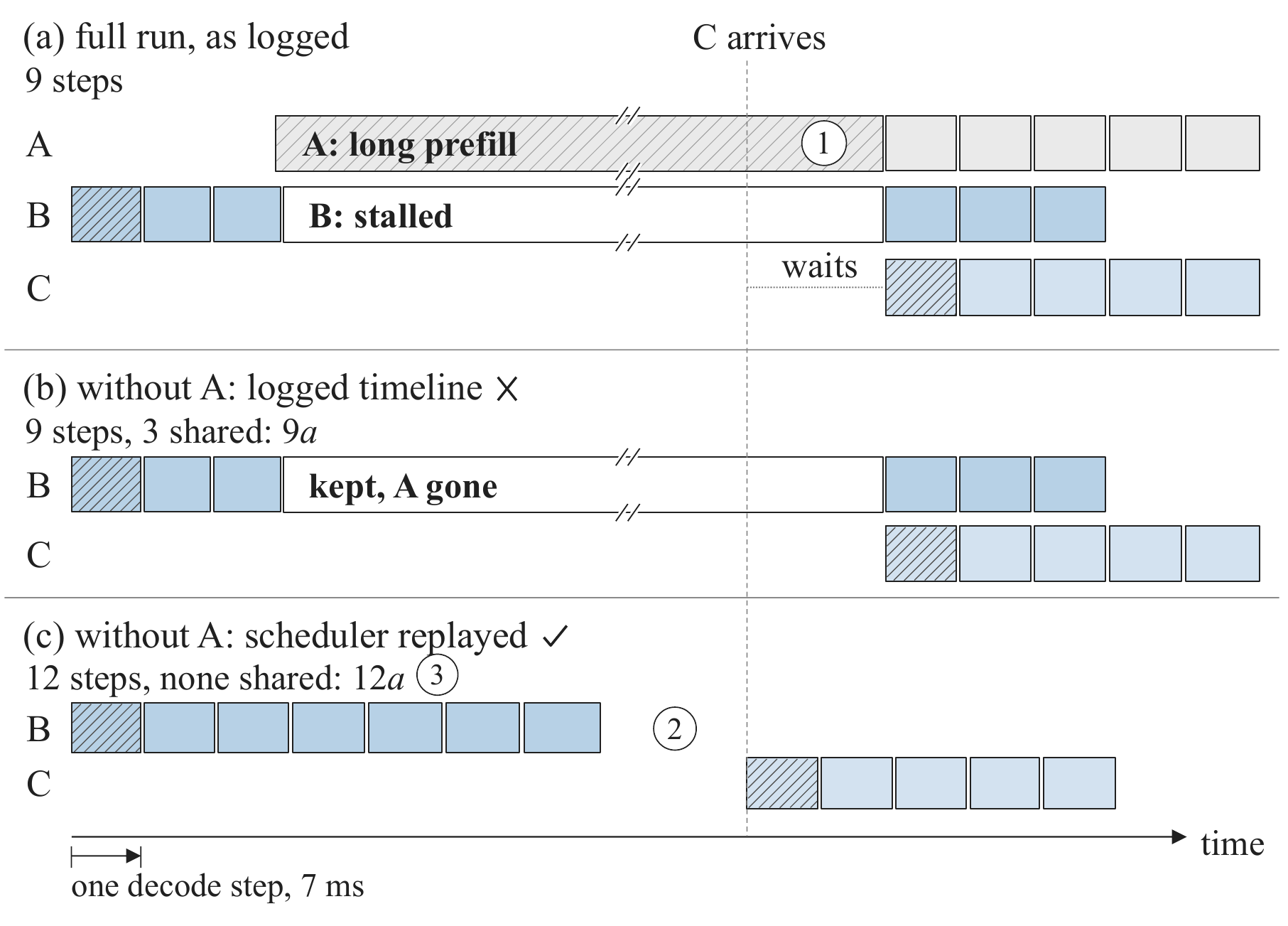}
\caption{Why the scheduler is replayed. (a)~In the full run, the long prefill of A \circnum{1} stalls B while C arrives. (b)~Read off the logged timeline, the coalition $\{B,C\}$ takes 9 steps, and 3 of them are shared. (c)~When replayed, B finishes before C arrives \circnum{2}, and none of the 12 steps is shared \circnum{3}.}
\label{fig:replay}
\end{figure}

As shown in Figure~\ref{fig:replay}, we reconstruct the steps of a coalition under continuous batching by replaying the scheduler on a CPU. The energy law still holds on the steps that the coalition goes through, so the game becomes
\begin{equation}
\hat v(S) \;=\; \sum_{i\in S} e_i \;+\; a\,\bigl|\mathcal{T}(S)\bigr|,
\label{eq:replaygame}
\end{equation}
where $\mathcal{T}(S)$ are the steps that the scheduler produces for $S$.

A CPU can replay the vLLM scheduler because it is deterministic given arrival times, input lengths, and output lengths. Each step serves the running requests first within the token budget and then admits arrived requests in arrival order into the remaining budget. Which requests have arrived by the next step depends on the duration of the current step, which we predict as $\hat t_k=\tau_0+\tau_1 P_k+\tau_2 n_k+\tau_3 L_k+\tau_4 A_k$, linear in the prefill tokens $P_k$, decoding requests $n_k$, decode context $L_k$ and prefill attention $A_k$ of step $k$. Fitted on the attributed run alone, this model has a median error of 0.2--4.8\% per step. Fitting it on 14 to 30 runs of the same workload instead changes the \Lone error of the shares by at most 0.005, so one run is enough. We reimplement the vLLM scheduler for the replay. For other engines, the scheduler code itself can be driven on a CPU, as GPU-free emulators do~\cite{agrawal2026revati}.

In a distributed deployment, a least-loaded router sends each arriving request to the node with the fewest requests in flight, so removing one request can move later ones to another node. We therefore also replay the router: at each arrival, it picks a node from the replayed state of the nodes. The energy of a coalition is Eq.~\eqref{eq:replaygame} summed over the nodes, each with the coefficients of its GPUs. Any router that decides deterministically from the state it observes, such as round-robin, can be replayed this way.

\subsection{Sampling the Correction}
\label{sec:sampling}

Under continuous batching, the steps change with the coalition, so we compute the Shapley value of the replay game of Eq.~\eqref{eq:replaygame} numerically. For 8 requests, we enumerate all 255 coalitions. Larger windows need permutation sampling~\cite{castro2009polynomial}, and sampling the replay game directly converges slowly (\S\ref{sec:eval:online}).

We therefore sample only the correction to the closed form $\shap_i$ of Eq.~\eqref{eq:rule}. Within one permutation, the marginal contributions of the model game are close to those of the replay game because the two games differ only where the schedule changes. Let $\Delta_i(\sigma)$ and $\Delta^{\log}_i(\sigma)$ be the marginal contributions of request $i$ in permutation $\sigma$ in the replay game and in the model game.

\begin{proposition}
\label{thm:cv}
For $Q$ permutations $\sigma_1,\dots,\sigma_Q$, each uniformly distributed,
\begin{equation}
\tilde\shap_i \;=\; \shap_i \;+\; \frac{1}{Q}\sum_{q=1}^{Q}\bigl[\Delta_i(\sigma_q)-\Delta^{\log}_i(\sigma_q)\bigr]
\label{eq:cv}
\end{equation}
is an unbiased estimate of the Shapley value of request $i$ in the replay game.
\end{proposition}

\begin{proof}
By Eq.~\eqref{eq:shapley}, the Shapley value of a game is the expected marginal contribution of $i$ over a uniformly random permutation. The expectation of $\Delta_i(\sigma_q)$ is therefore the Shapley value in the replay game, and that of $\Delta^{\log}_i(\sigma_q)$ is $\shap_i$, so $\shap_i$ and the mean of $\Delta^{\log}_i(\sigma_q)$ cancel in expectation.
\end{proof}

This estimate is a control variate, and its variance comes only from the difference between the two games. On one engine, it therefore needs far fewer permutations than sampling the replay game itself (\S\ref{sec:eval:online}). In the distributed deployment, once the routing changes, the model game with the logged routing no longer tracks the replay, and the control variate helps little. The permutations run in parallel, in at most 105\,MiB each.

\subsection{Online Ledger}
\label{sec:ledger}

The online ledger wraps three methods of the vLLM scheduler and needs no change to the engine. After each scheduling decision, the ledger adds to every scheduled request its term $e_{i,k}$ of Eq.~\eqref{eq:law} and its part $a/m_k$ of the shared energy. A background thread closes an attribution window every second at a counter update. It subtracts the idle power over the parts of the window that no step covers and over host stalls, so only waiting for arrivals is left out. It then divides the remaining energy among the requests in proportion to the parts they accumulated in the window, splitting a step that spans two windows by its overlap with each.

\section{Evaluation}
\label{sec:eval}

Our evaluation answers four questions. How close is \approach to the exact Shapley value of 8-request batches on three GPU generations (\S\ref{sec:eval:exact})? Does it stay close in static batches of up to 256 requests and for up to 128 requests under continuous batching (\S\ref{sec:eval:large}, \S\ref{sec:eval:serving})? Does it carry over to a distributed deployment (\S\ref{sec:eval:distributed}), to MoE models (\S\ref{sec:eval:moe}) and to prefix caching (\S\ref{sec:eval:prefix})? And what do the online ledger and the replay cost (\S\ref{sec:eval:online})?

\subsection{Setup}
\label{sec:setup}\label{app:details}\label{app:scale}

All experiments use the setup of \S\ref{sec:platform}, with asynchronous scheduling disabled so that each step is scheduled after the previous one finishes. Prefix caching is disabled except in \S\ref{sec:eval:prefix}.

\textbf{Measured Shapley reference.} We compare every allocation with Shapley values measured on the GPU (\S\ref{sec:game}). For 8 requests, we run all 255 subsets of a request group in random order so that thermal drift is not a systematic error, and compute Eq.~\eqref{eq:shapley} exactly. Under continuous batching, the requests of a group arrive 0.5\,s apart, and a subset keeps the arrival times of its members. A budget of 65{,}536 tokens per step, or 262{,}144 for large static batches, prefills every prompt in one step. For $N\ge16$, we estimate the Shapley value by permutation sampling~\cite{castro2009polynomial} with 50--150 antithetic pairs. We run all $N-1$ proper prefixes of each sampled order and its reverse. We call the two references the \emph{exact} and the \emph{sampled} Shapley value. Together they take 529{,}513 runs, and each run is checked for preemptions and exact output lengths. The busy energy leaves out host stalls, which are the part of a step beyond three times its expected duration. The measurement follows~\citet{luo2026request}.

\textbf{Metrics.} The error of an allocation is its normalized \Lone distance to the reference (\S\ref{sec:example}), with both adding up to the measured energy $v(\mathcal{N})$ of the full run. It is 0 when the allocation matches the reference and 2 when the two give all energy to different requests. For $N\ge16$, the sampled Shapley value $\hat\shap$ has a sampling error that depends on the measured permutations, so we compare every allocation with a reference that cancels most of it. On the same permutations, the game $M$ of \approach gives the estimate $\hat\shap^{M}$. Its Shapley value $\shap^{M}$ comes in closed form under static batching and from 2{,}000 CPU permutation pairs under continuous batching (3{,}000 in the distributed deployment). The measured and the modeled marginal contributions move together, so $\hat\shap^{M}-\shap^{M}$ shows how far the measured permutations are off, and the reference $r=\hat\shap-(\hat\shap^{M}-\shap^{M})$ removes it. As in Proposition~\ref{thm:cv}, $r$ is unbiased whatever $M$ is, and a less accurate $M$ only makes it noisier. Against $r$, the error of \approach is the difference between its game and the measured game on the same permutations (\emph{paired}), plus the sampling error of its CPU estimate, in expectation over the CPU pairs. The reference does not favor \approach. For request $i$, $\shap^{M}_i-r_i=\hat\shap^{M}_i-\hat\shap_i$, whose expectation is $\shap^{M}_i-\shap_i$, the error of \approach against the Shapley value $\shap_i$ of the measured game, and the sampling noise of this difference can only raise its \Lone.

\textbf{Noise.} Identical runs on the H100 differ in energy by 0.1--4.3\%, most for short runs. A marginal contribution is the difference of two runs, so the reference is noisy, and we read every result against a noise level. The \emph{reference noise} applies to every allocation. For 8 requests, it is the \Lone between two exact Shapley values from measuring all 255 subsets twice, and an allocation equal to the true value would be expected at $1/\sqrt{2}$ of it. For $N\ge16$, it is the mean \Lone between bootstrap resamples of $r$, over the measured and the CPU pairs, and $r$ itself. Because the error of \approach against $r$ is paired, it can also be read against the \emph{paired noise}, the \Lone that the paired comparison would have if the game of \approach matched the measured game. It is $\sqrt{2/\pi}\sum_i \mathrm{se}_i/v(\mathcal{N})$, with $\mathrm{se}_i$ the standard error of the paired difference of request~$i$.

\textbf{Baselines.} Token, Solo and JCalib are defined in \S\ref{sec:example}. Where Solo has no run of a request alone, it uses the runs in which the request came first in a sampled permutation. JCalib is trained on the Shapley values of other request sets of the same workload, exact or sampled, whichever makes it most accurate.

\subsection{Exact Shapley Values in 8-Request Batches}
\label{sec:eval:exact}

\begin{table*}[t]
\caption{Mean \Lone over request groups against the exact Shapley value of 8 requests. Noise is the reference noise, the \Lone between two measurements of the same exact Shapley values.}
\label{tab:main}
\centering
\begin{small}
\setlength{\tabcolsep}{2.2pt}
\begin{tabular}{llccc@{\hskip 0.08in}ccccc@{\hskip 0.08in}ccccc}
\toprule
\multirow{2}{*}{\hd{GPU}} & \multirow{2}{*}{\hd{Model}} & \multirow{2}{*}{\hd{Arch.}} & \multirow{2}{*}{\hd{GPUs}} & \multirow{2}{*}{\hd{Groups}} & \multicolumn{5}{c}{\hd{Static batching}} & \multicolumn{5}{c}{\hd{Continuous batching}} \\
\cmidrule(lr){6-10}\cmidrule(lr){11-15}
 & & & & & \hd{Noise} & \hd{Token} & \hd{Solo} & \hd{JCalib} & \hd{\approach} & \hd{Noise} & \hd{Token} & \hd{Solo} & \hd{JCalib} & \hd{\approach} \\
\midrule
\multirow{10}{*}{\shortstack[l]{H100\\(Hopper)}} & Qwen2.5-7B & dense & 1 & 16 & 0.009 & 0.559 & 0.274 & 0.175 & \textbf{0.008} & 0.007 & 0.516 & 0.140 & 0.159 & \textbf{0.011} \\
 & Qwen2.5-7B & dense & 2 & 4 & 0.011 & 0.594 & 0.139 & 0.091 & \textbf{0.009} & 0.004 & 0.565 & 0.142 & 0.151 & \textbf{0.019} \\
 & Qwen2.5-14B & dense & 1 & 16 & 0.006 & 0.551 & 0.269 & 0.170 & \textbf{0.012} & 0.005 & 0.521 & 0.155 & 0.138 & \textbf{0.014} \\
 & Qwen2.5-14B & dense & 2 & 8 & 0.006 & 0.584 & 0.254 & 0.152 & \textbf{0.011} & 0.009 & 0.552 & 0.158 & 0.131 & \textbf{0.018} \\
 & Qwen2.5-14B & dense & 4 & 8 & 0.003 & 0.586 & 0.254 & 0.153 & \textbf{0.011} & 0.006 & 0.550 & 0.137 & 0.141 & \textbf{0.013} \\
 & Qwen2.5-72B & dense & 4 & 4 & 0.006 & 0.567 & 0.130 & 0.094 & \textbf{0.009} & 0.006 & 0.569 & 0.166 & 0.097 & \textbf{0.011} \\
 & Gemma-4-12B & dense & 1 & 4 & 0.004 & 0.585 & 0.127 & 0.097 & \textbf{0.021} & 0.006 & 0.598 & 0.177 & 0.118 & \textbf{0.019} \\
 & Qwen3.8-27B & hybrid & 1 & 4 & 0.002 & 0.582 & 0.128 & 0.099 & \textbf{0.007} & 0.003 & 0.561 & 0.144 & 0.092 & \textbf{0.013} \\
 & gpt-oss-20b & MoE & 1 & 4 & 0.027 & 0.438 & 0.091 & 0.169 & \textbf{0.015} & 0.009 & 0.452 & 0.118 & 0.173 & \textbf{0.020} \\
 & Qwen3-30B-A3B & MoE & 1 & 4 & 0.006 & 0.532 & 0.079 & 0.086 & \textbf{0.011} & 0.006 & 0.523 & 0.089 & 0.090 & \textbf{0.040} \\
\midrule
\multirow{6}{*}{\shortstack[l]{A800\\(Ampere)}} & Qwen2.5-7B & dense & 1 & 8 & 0.016 & 0.595 & 0.226 & 0.116 & \textbf{0.007} & 0.012 & 0.578 & 0.190 & 0.141 & \textbf{0.015} \\
 & Qwen2.5-72B & dense & 4 & 4 & 0.005 & 0.547 & 0.124 & 0.090 & \textbf{0.007} & 0.004 & 0.530 & 0.119 & 0.086 & \textbf{0.013} \\
 & Gemma-4-12B & dense & 1 & 4 & 0.003 & 0.539 & 0.127 & 0.092 & \textbf{0.015} & 0.006 & 0.531 & 0.139 & 0.090 & \textbf{0.018} \\
 & Qwen3.8-27B & hybrid & 1 & 4 & 0.003 & 0.557 & 0.127 & 0.097 & \textbf{0.008} & 0.003 & 0.542 & 0.115 & 0.089 & \textbf{0.012} \\
 & gpt-oss-20b & MoE & 1 & 4 & 0.008 & 0.411 & 0.085 & 0.198 & \textbf{0.020} & 0.011 & 0.431 & 0.127 & 0.185 & \textbf{0.026} \\
 & Qwen3-30B-A3B & MoE & 1 & 4 & 0.015 & 0.525 & 0.081 & 0.087 & \textbf{0.011} & 0.004 & 0.506 & 0.098 & 0.081 & \textbf{0.029} \\
\midrule
\multirow{5}{*}{\shortstack[l]{PRO 6000\\(Blackwell)}} & Qwen2.5-7B & dense & 1 & 8 & 0.007 & 0.613 & 0.242 & 0.131 & \textbf{0.008} & 0.006 & 0.590 & 0.195 & 0.150 & \textbf{0.014} \\
 & Gemma-4-12B & dense & 1 & 4 & 0.009 & 0.555 & 0.134 & 0.089 & \textbf{0.011} & 0.005 & 0.544 & 0.157 & 0.091 & \textbf{0.014} \\
 & Qwen3.8-27B & hybrid & 1 & 4 & 0.002 & 0.588 & 0.137 & 0.105 & \textbf{0.004} & 0.003 & 0.567 & 0.127 & 0.104 & \textbf{0.005} \\
 & gpt-oss-20b & MoE & 1 & 4 & 0.027 & 0.463 & 0.091 & 0.149 & \textbf{0.023} & 0.052 & 0.471 & 0.174 & 0.180 & \textbf{0.035} \\
 & Qwen3-30B-A3B & MoE & 1 & 4 & 0.005 & 0.520 & 0.072 & 0.095 & \textbf{0.013} & 0.013 & 0.514 & 0.099 & 0.080 & \textbf{0.024} \\
\bottomrule
\end{tabular}
\end{small}
\end{table*}

Table~\ref{tab:main} compares \approach with the exact Shapley value for every platform and model. The two main models on the H100 have four groups per trace, and the other rows have one or two.

\textbf{Static batching.} Under static batching, \approach stays close to the reference noise. Here every coalition goes through the logged steps that hold its members (\S\ref{sec:theory}), so the comparison tests the energy law and Proposition~\ref{thm:main}. Over all rows, \approach is at 0.004--0.023 against reference noise of 0.002--0.027, and averaged over the 124 groups it is at 0.011 against 0.008. Token is at 0.411--0.613, $20\times$ to $147\times$ further than \approach. Solo is at 0.072--0.274, $4.0\times$ to $34\times$ further. JCalib is trained on exact Shapley values and is at 0.086--0.198. \approach uses no such values, yet its error is lower than that of JCalib in all 124 groups, and $11\times$ lower in the median group.

\textbf{Continuous batching.} For the dense and hybrid models, \approach is at 0.005--0.019 against reference noise of 0.003--0.012, and at 0.014--0.033 without replay. It is within 0.003 of what the timelines of the coalitions themselves give, which is what an exact replay would give. Averaged over all 124 groups, \approach is at 0.017 against a reference noise of 0.008. Token is at 0.431--0.598, $13\times$ to $113\times$ further than \approach. Solo is at 0.089--0.195, $2.2\times$ to $25\times$ further. Even without replay, the error of \rulename is lower than that of JCalib in 122 of the 124 groups, and $5.4\times$ lower in the median group.

\textbf{Multiple GPUs.} \approach is as accurate with tensor parallelism as on one GPU. Over two and four H100s, it is at 0.009--0.011 under static batching, against 0.008--0.011 for the same groups on one GPU, and at 0.011--0.019 under continuous batching. Qwen2.5-72B on four A800s is at 0.007 and 0.013.

\textbf{Per request.} The \Lone is the energy-weighted mean of the per-request relative error $|\hat\shap_i-\shap_i|/\shap_i$. Over all requests of the static groups, \approach has a median relative error of 1.1\%, and a p95 of 5.6\% over the requests above 1\% of the energy of their batch. Under continuous batching, these are 1.5\% and 8.9\%. The low \Lone therefore does not hide large errors on single requests.

\subsection{Static Batching with up to 256 Requests}
\label{sec:eval:large}

\begin{figure}[h]
\centering
\includegraphics[width=0.92\linewidth]{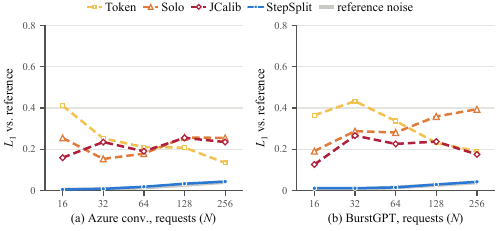}
\caption{\Lone against the reference under static batching with up to 256 requests.}
\label{fig:scalestatic}
\end{figure}

Figure~\ref{fig:scalestatic} compares \approach and the baselines with the reference $r$ of \S\ref{sec:setup} for Qwen2.5-7B on the H100. A batch of $N$ requests holds the first $N$ of 256 requests drawn from the Azure conversation or the BurstGPT trace. Each trace and size has 100--150 antithetic permutation pairs. Because requests finish at different steps, a step holds 7 to 102 requests on average. At 16 and 32 requests, \approach is at 0.005--0.011 against a reference noise of 0.004--0.007. From 64 to 256 requests, its error grows to 0.015--0.043, but the reference noise grows with it, to 0.014--0.041. At every size, \approach thus agrees with the measured marginal contributions within their noise. The reference takes 2{,}484 to 51{,}008 GPU runs per trace and size, while \approach uses one run. The baselines stay far from the reference at every size. Token is at 0.134--0.433 and Solo is at 0.153--0.394. JCalib is at 0.126--0.266.

\subsection{Continuous Batching with up to 128 Requests}
\label{sec:eval:serving}

\begin{figure}[h]
\centering
\includegraphics[width=0.92\linewidth]{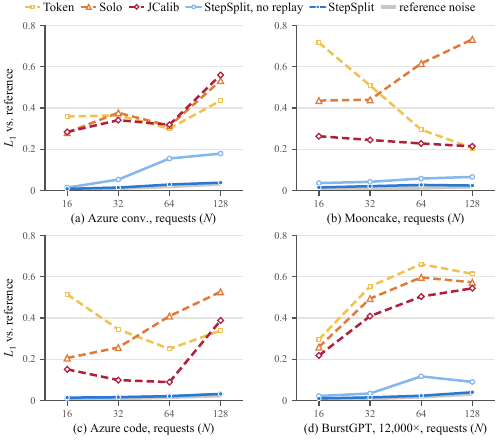}
\caption{\Lone against the reference in trace windows of up to 128 requests.}
\label{fig:scaleserving}
\end{figure}

\textbf{Setup.} We run Qwen2.5-7B on the H100 under continuous batching on windows of four traces. Each window holds the first $N$ requests of a 128-request window and has 50 antithetic permutation pairs. To reach larger batches, the Azure conversation, Mooncake and Azure code windows run at $10\times$ the trace rate. The BurstGPT arrivals are compressed $12{,}000\times$, which matches the request rate of the Azure conversation windows. Unlike the 8-request groups, requests arrive as in the trace, with 2.8 to 36.4 requests per step. In addition, the default budget of 16{,}384 tokens per step splits long inputs over several steps while new requests wait. As in the 8-request groups, a subset keeps the arrival times of its members, as if the other requests had never been sent.

\textbf{Without replay, the error grows with the window.} As shown in Figure~\ref{fig:scaleserving}, the error without replay on the Azure conversation trace grows with the window, from 0.014 at 16 requests to 0.179 at 128. This error is above the reference noise and has a consistent direction. A long prefill stalls the batch while arrivals accumulate. Afterwards more requests decode together, so the same tokens take fewer steps. The logged timeline therefore over-charges long inputs and under-charges long outputs. Its size differs across traces. On BurstGPT, at the same request rate, it reaches 0.117 at 64 requests. On the Mooncake trace, it is 0.036--0.066. There, 121 of 128 requests wait in the queue, and removing requests mostly shortens the wait. The Azure code trace has 2.8 to 10.0 requests per step, so removing a request barely moves the others, and the logged timeline is within 0.004 of the replay.

\textbf{\approach removes most of this error.} On the Azure conversation trace, \approach is at 0.008--0.038, $3.7\times$ to $5.3\times$ lower than without replay from 32 requests on, and on BurstGPT at 0.010--0.040, $2.3\times$ to $5.1\times$ lower. It is 0.015--0.027 on the Mooncake trace and 0.014--0.032 on Azure code. This is $1.3\times$ to $2.0\times$ the paired noise on three traces and $1.3\times$ to $3.8\times$ on Mooncake. The gain comes from the replay, which reproduces the engine. Its full-window step counts are within two steps of the logged ones, and its subset step counts are within 0.7\% of those on the timelines of the coalitions, against 0.6--10\% for the logged timeline. The baselines are far from the reference in every window. Token is at 0.206--0.717, Solo at 0.206--0.733 and JCalib at 0.089--0.560. Even without replay, \approach is lower than all three in every window.

\subsection{Distributed Deployment}
\label{sec:eval:distributed}

\begin{table}[ht]
\caption{\Lone in the distributed deployment. Noise is the reference noise.}
\label{tab:distributed}
\centering
\begin{footnotesize}
\setlength{\tabcolsep}{3.2pt}
\begin{tabular}{lcccccc}
\toprule
\hd{Trace} & \hd{$N$} & \hd{Noise} & \hd{Token} & \hd{Solo} & \hd{JCalib} & \hd{\approach} \\
\midrule
\multirow{3}{*}{Azure conv.} & 32 & 0.020 & 0.431 & 0.319 & 0.300 & \textbf{0.020} \\
 & 64 & 0.037 & 0.342 & 0.298 & 0.303 & \textbf{0.038} \\
 & 128 & 0.054 & 0.405 & 0.466 & 0.468 & \textbf{0.064} \\
\midrule
\multirow{3}{*}{Mooncake} & 32 & 0.019 & 0.632 & 0.294 & 0.140 & \textbf{0.029} \\
 & 64 & 0.025 & 0.402 & 0.485 & 0.216 & \textbf{0.028} \\
 & 128 & 0.035 & 0.265 & 0.651 & 0.389 & \textbf{0.038} \\
\bottomrule
\end{tabular}
\end{footnotesize}
\end{table}

\textbf{Setup.} We use the distributed deployment of \S\ref{sec:platform}. Its router sends each request at its arrival time to the node with the fewest requests in flight, like the vLLM load balancer. Both nodes use the coefficients of Qwen2.5-72B on four H100s. The windows hold 32, 64 and 128 requests of the Azure conversation and Mooncake traces at $10\times$ their rate. Each window has 50 antithetic permutation pairs.

\textbf{The routing changes with the subset.} Removing requests changes which node the router sends the remaining requests to. On 40 random subsets of each window, the energy law predicts the energy of a subset with an average error of 0.17--0.42\% when given the steps that the subset really takes. Given the logged steps of the full run, the error is 1.9--6.0\%, and given the steps of the replay, it is 0.20--0.63\%.

\textbf{\approach stays close to the reference noise.} \approach is at 0.020--0.064 against reference noise of 0.019--0.054 (Table~\ref{tab:distributed}). The replay is also robust to near-ties. Where near-\hspace{0pt}simultaneous arrivals and completions flip 0.7\% of the routing decisions, \approach stays at 0.038 against reference noise of 0.035. Replaying only the schedulers, with every request kept on its node of the full run, gives 0.070--0.178, against 0.074--0.180 without replay, so the router must be replayed as well. None of the baselines comes close. Token is at 0.265--0.632, Solo at 0.294--0.651 and JCalib at 0.140--0.468.

\subsection{Mixture-of-Experts Models}
\label{sec:eval:moe}

\begin{table}[ht]
\caption{\Lone of the split per expert (\S\ref{sec:eval:moe}). Noise is the paired noise, and Equal splits the shared energy of each step equally among its requests, whatever experts they use.}
\label{tab:moe}
\centering
\begin{footnotesize}
\setlength{\tabcolsep}{2.6pt}
\begin{tabular}{lcccc@{\hskip 0.08in}ccc}
\toprule
\multirow{2}{*}{\hd{Trace}} & \multirow{2}{*}{\hd{$N$}} & \multicolumn{3}{c}{\hd{Qwen3-30B-A3B}} & \multicolumn{3}{c}{\hd{gpt-oss-20b}} \\
\cmidrule(lr){3-5}\cmidrule(lr){6-8}
 & & \hd{Noise} & \hd{Equal} & \hd{Expert} & \hd{Noise} & \hd{Equal} & \hd{Expert} \\
\midrule
\multirow{3}{*}{Azure conv.} & 16 & 0.007 & 0.031 & \textbf{0.016} & 0.011 & 0.034 & \textbf{0.023} \\
 & 32 & 0.011 & 0.060 & \textbf{0.022} & 0.013 & 0.053 & \textbf{0.016} \\
 & 64 & 0.028 & 0.065 & \textbf{0.036} & 0.028 & 0.043 & \textbf{0.035} \\
\midrule
\multirow{3}{*}{Mooncake} & 16 & 0.006 & 0.011 & \textbf{0.007} & 0.008 & 0.030 & \textbf{0.013} \\
 & 32 & 0.009 & 0.018 & \textbf{0.015} & 0.020 & 0.036 & \textbf{0.032} \\
 & 64 & 0.013 & 0.033 & \textbf{0.012} & 0.022 & 0.031 & \textbf{0.026} \\
\bottomrule
\end{tabular}
\end{footnotesize}
\end{table}

\textbf{The shared energy grows with the requests in a step.} A step of a mixture-of-experts (MoE) model reads only the experts its tokens are routed to, and more requests touch more experts. On the H100, from 1 to 8 requests per step, the shared energy grows from 1.3 to 2.4\J for gpt-oss-20b and from 2.5 to 5.9\J for Qwen3-30B-A3B on the two GPUs of its serving windows, but by 4\% for the dense Qwen2.5-7B. The shared term becomes a function $a(m)$, which the microbenchmark measures, and Proposition~\ref{thm:main} gives each request $a(m_k)/m_k$ of every step it is in. The MoE rows of Table~\ref{tab:main} use this split and are at 0.020--0.040 under continuous batching.

\textbf{The read of each expert should be split among its users.} In an MoE step, the weights outside the experts are read for all requests in the step, but an expert is read only for the requests routed to it. Here $a_0(m_k)$ is the shared energy of the weights outside the experts, and $w$ is the energy of reading all $X$ experts of the MoE layers once. With $R_{i,k}$ as the experts that request $i$ uses in step $k$ and $q_{k,x}$ as the number of requests in step $k$ routed to expert $x$, the Shapley value is
\begin{equation}
\shap_i \;=\; e_i + \sum_{k:\,i\in M_k}\Bigl[\frac{a_0(m_k)}{m_k} + \frac{w}{X}\sum_{x\in R_{i,k}}\frac{1}{q_{k,x}}\Bigr].
\label{eq:expert}
\end{equation}
Each expert is a game of the form of Proposition~\ref{thm:main}, which is the multicast cost sharing of~\citet{feigenbaum2001sharing} with experts in place of links. The rule needs the routing log of the run and assumes that a request is routed to the same experts in every coalition.

\textbf{Splitting per expert is more accurate than splitting each step equally.} Table~\ref{tab:moe} evaluates the split per expert on the H100. The windows hold 16, 32 and 64 requests of the Azure conversation and Mooncake traces at $10\times$ the trace rate, with 50 permutation pairs each and 3.3 to 28 requests per step. To compare only the two splits, we use the steps that each subset really takes instead of the replay, and fit the law, including $a_0$ and $w$, on the runs of the window instead of the microbenchmark. In all 12 windows, the split per expert has a lower error than splitting each step equally, at 0.007--0.036 against 0.011--0.065.

\subsection{Prefix Caching}
\label{sec:eval:prefix}

\begin{table}[ht]
\caption{\Lone with prefix caching. Noise is the reference noise, and First charges a block to the request that computed it.}
\label{tab:prefix}
\centering
\begin{footnotesize}
\setlength{\tabcolsep}{3.0pt}
\begin{tabular}{llcccccc}
\toprule
\hd{Batching} & \hd{Group} & \hd{Noise} & \hd{Token} & \hd{Solo} & \hd{JCalib} & \hd{First} & \hd{\approach} \\
\midrule
\multirow{4}{*}{Static} & Mooncake A & 0.004 & 1.307 & 0.197 & 0.156 & 0.094 & \textbf{0.005} \\
 & Mooncake B & 0.005 & 1.368 & 0.181 & 0.087 & 0.132 & \textbf{0.005} \\
 & System prompt & 0.008 & 0.538 & 0.133 & 0.112 & 0.020 & \textbf{0.002} \\
 & Nested & 0.006 & 0.659 & 0.169 & 0.137 & 0.039 & \textbf{0.003} \\
\midrule
\multirow{4}{*}{Continuous} & Mooncake A & 0.007 & 1.240 & 0.157 & 0.133 & 0.091 & \textbf{0.010} \\
 & Mooncake B & 0.007 & 1.417 & 0.243 & 0.143 & 0.090 & \textbf{0.011} \\
 & System prompt & 0.008 & 0.616 & 0.327 & 0.305 & 0.027 & \textbf{0.004} \\
 & Nested & 0.003 & 0.712 & 0.351 & 0.361 & 0.029 & \textbf{0.008} \\
\bottomrule
\end{tabular}
\end{footnotesize}
\end{table}

\textbf{A shared prefix should be split equally.} With prefix caching, when several requests share a prompt prefix, the engine computes its prefill once for the first of them, and the others reuse it. This prefill is a second shared cost like the weight read, because a block is computed once whenever one of the requests that contain it is present. The Shapley value therefore splits the prefill energy of each block equally among the requests that contain it, and \approach adds this split to \rulename. In the replay, a block is cached for a request once a request admitted earlier has computed it.

\textbf{Setup.} We enable prefix caching, empty the cache before every run, and measure all 255 subsets of four groups of 8 requests with Qwen2.5-7B on the H100. Mooncake A and B come from the Mooncake trace, which records the prompt blocks of every request. In System prompt, 8 conversation prompts follow a common 1{,}024-token system prompt. In Nested, each prompt is a prefix of the next, as in a multi-turn conversation. The cache saves 5--25\% of the group energy. We compare \approach with \emph{First}, which charges each block to the request that computed it, as reading the steps of the run does. Both use coefficients from the microbenchmark, which never saw a cached prefix. JCalib is trained on the other three groups or on the 8-request groups of the same trace, whichever is more accurate, and Token bills cached tokens as if they were computed.

\textbf{With the cache on, \approach stays close to the reference noise.} As shown in Table~\ref{tab:prefix}, \approach is at 0.002--0.011 against reference noise of 0.003--0.008. In contrast, First is at 0.020--0.132, Token at 0.538--1.417, Solo at 0.133--0.351 and JCalib at 0.087--0.361. On the Mooncake trace windows of 16 and 32 requests with the cache on, the cached tokens match the prefix tree in every run, and the microbenchmark law with the prefill terms of the tree predicts the energy of a subset with an average error of 0.7--0.9\%.

\subsection{Online Ledger and Replay Cost}
\label{sec:eval:online}

We run the ledger, which wraps the vLLM scheduler without changing the engine, on the H100 with Qwen2.5-7B on two 32-request trace windows at $3\times$ the trace rate. To measure its overhead, we also run it on a 128-request window with the ledger off and on in alternation. Its shares equal the closed form recomputed from the scheduler log and sum to the busy energy of the counter within 0.2\%. Attribution takes 47\,$\mu$s per step, which is 0.7\% of the mean step time. The ledger changes completion time, throughput and time per output token by at most 0.3\%, which is within run-to-run variation.

Against the reference of the two windows, the ledger is at 0.086 and 0.114. The closed form on the logged timeline of the measured full run is at 0.076 and 0.112, so settling every second on the timeline of a separate run adds only 0.010 and 0.002. Because the ledger settles on the logged timeline, it carries the error of the logged timeline (\S\ref{sec:eval:serving}). A provider can therefore bill while the engine runs and settle the bill after the run by replay, which is at an \Lone of 0.008--0.040 on the trace windows of \S\ref{sec:eval:serving}.

\begin{figure}[ht]
\centering
\includegraphics[width=0.92\linewidth]{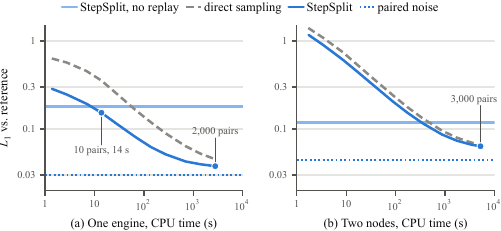}
\caption{\Lone against the CPU time of the replay on the Azure conv. window of 128 requests.}
\label{fig:cpubudget}
\end{figure}

\textbf{The replay finishes in minutes.} On one core of an Intel Xeon 6530P, one replay of a window of 16 to 128 requests takes 0.23--7.74\,ms on average, and the 255 subsets of an 8-request group take 0.02--0.69\,s. For larger windows, \approach samples permutation pairs (\S\ref{sec:sampling}) and replays every prefix of each, so its cost grows with the number of pairs. As shown in Figure~\ref{fig:cpubudget}, the error of \approach falls as more pairs are run, so a provider can spend more CPU time for a more accurate bill. On the Azure conversation window of 128 requests, which takes the longest to replay, \approach falls below the logged timeline within 10 pairs, or 14 CPU seconds. Over the trace windows of \S\ref{sec:eval:serving}, an estimate within an \Lone of 0.010 of its limit takes 1--2{,}446 pairs, which is $4.3\times$ to $870\times$ fewer than sampling the replay game directly, and at most 56.5 CPU minutes. In the distributed deployment, the 3{,}000 pairs of Table~\ref{tab:distributed} take 86.8 CPU minutes on the same window, and \approach is at 0.064 against 0.058 in the limit. The pairs are independent and run in parallel, so on 32 cores both take under 3 minutes.

\section{Discussion}
\label{sec:discussion}

\textbf{Deployment choices set the shared term.} Per-request shares make the energy cost of these choices visible. Tensor parallelism over two and four H100s makes a step $1.4\times$ to $2.2\times$ faster, but it raises the shared term by 14--40\% and the energy of the same work by 18--40\%. Because this choice acts on the shared term, a provider can charge the premium of a tensor-parallel deployment to the requests that need its latency. Similarly, the shared energy of an MoE step grows with the experts that its tokens touch (\S\ref{sec:eval:moe}). An MoE engine that groups requests routed to the same experts would therefore lower the shared energy of their steps.

\textbf{Bills of users and of a continuous service.} Because the engine schedules individual requests, the bill of a user is the sum of the shares of the requests from that user. In a step of $m_k$ requests, a user with $n$ of them pays $n\,a/m_k$, whereas a game with users as players would give every user in the step the same part. In addition, the share of a request depends on the requests it ran with. A provider that needs a price before the run can therefore quote one and use the fair share to settle the bill or to report the energy after the run. Finally, a provider attributes a continuous service window by window. Each window is its own game. Its players are the requests that arrive in it, and the state of the engine at its start, such as requests still running and cached prefixes, is held fixed. The requests of a window therefore pay the energy that they add to this state, and the requests still running pay in the window in which they arrived.

\textbf{The carbon of a request.} Providers also report the carbon footprint of each request~\cite{elsworth2025google,rince2025ecologits}. The carbon of a step is its energy times the carbon intensity of the grid, which changes far more slowly than a step. Because the Shapley value is linear in the game, the carbon share of a request in a window of constant intensity is its energy share times that intensity, and its footprint is the sum over the windows it spans. The online ledger can therefore report the carbon of each request as it bills its energy, and the replay settles both after the run.

\section{Related Work}
\label{sec:related}

\textbf{The Shapley value in computer architecture.} \citet{choi2020fairwrite} attribute the writes of a shared flash cache to the consolidated workloads and model the subsets they do not observe. Concorde~\cite{nasresfahany2025concorde} divides the performance difference between two CPU designs among microarchitectural components, and ShapleyIQ~\cite{li2024shapleyiq} reduces the exponential cost of the Shapley value in a microservice system by decomposition. Like these works, we model the subsets we do not run and decompose the game into one game per step.

\textbf{Energy attribution.} Time sharing~\cite{islam2016new,sharma2024faasmeter,shen2013power} and Fair-CO2~\cite{han2025fairco2} split shared power or carbon by time rather than by steps. \citet{li2025fair} show that token-based rules violate the Shapley axioms. Neuralwatt bills each request for the server energy measured while it is in flight, in proportion to its share of the tokens of all requests then in flight~\cite{neuralwatt2026methodology}. In our measurements on 8-request groups, this rule has an \Lone error of 19.9--36.3\%, depending on whether output tokens are weighted as in Token, against 0.8--1.1\% for \approach. \citet{roy2026seam} attributes energy offline per step by activity weights, and \citet{yuan2026wattsonllm} approximate the Shapley value online. \citet{roy2026identifiability} argues that exact shares of the weight read need a replay of every subset, which our step game avoids. In contrast, \approach derives the Shapley value from the measured energy of a step and stays close to Shapley values measured on the GPU for up to 256 requests.

\textbf{Energy models and scheduler simulation.} Prior work characterizes inference energy~\cite{patel2024characterizing,patel2024splitwise}, models it for mostly homogeneous batches~\cite{wilkins2024offline}, or expresses it as a fixed cost per batch or step plus per-token costs~\cite{vellaisamy2026characterization,fortuna2026where}. Attribution needs such a law for arbitrary heterogeneous subsets, and our microbenchmark calibrates one once per model and GPU configuration. Revati~\cite{agrawal2026revati} runs serving schedulers without GPUs for capacity planning, whereas we replay the scheduler to find the steps of each subset. The cost of an MoE layer depends on the experts that a batch activates~\cite{gupta2024lynx}, and we split it by the routing.

\section{Conclusion}
\label{sec:conclusion}

Splitting the energy of a batch by tokens is far from the fair split, and the Shapley value has needed a GPU run for every subset of the requests. We showed by measurement that each step of batched inference costs a shared energy, mostly to read the weights, plus the energy that each request adds on its own. Under this structure, the Shapley value has a closed form computed from one run, and a CPU replay of the scheduler and the router supplies the steps when they change with the subset. Against Shapley values measured on the GPU, \approach is far closer than splitting by tokens or a regression trained on measured Shapley values, and it also runs online inside vLLM. We will release \approach, including the online ledger.

\bibliographystyle{ACM-Reference-Format}
\bibliography{refs}

\end{document}